\documentclass[a4paper,11pt]{article}
\usepackage{pos}

\usepackage{url}
\usepackage{nicefrac}
\usepackage{cuted}
\usepackage{slashed}
\usepackage{orcidlink}

\newcommand{\quark}{\mathfrak{q}}

\newcommand{\amp}{\mathcal{A}} 

\newcommand{\cffH}{\mathcal{H}}

\newcommand{\bq}{\bar{q}}
\newcommand{\bp}{\bar{p}}

\newcommand{\lop}{\mathscr{O}}

\newcommand{\PhiPlusF}{\Phi^{(+)}_f(\beta,\alpha,t)}
\newcommand{\PhiPlus}{\Phi^{(+)}(\beta,\alpha,t)}
\newcommand{\phiplus}{\Phi^{(+)}}

\newcommand{\hplus}{H^{(+)}}

\newcommand{\scale}{\mathbb{Q}}

\newcommand{\bu}{{\bar{u}}}

\newcommand{\dxi}{\partial_{\xi}}

\newcommand\lt[1]{{\left[#1\right]_{\rm LT}}} 

\definecolor{softGray}{rgb}{0.75, 0.75, 0.75}
\definecolor{darkred}{rgb}{0.55, 0.0, 0.0}

\title{Double DVCS amplitudes including kinematic twist-3 and 4 corrections}

\author*[a]{V.~Mart\'inez-Fern\'andez\,\orcidlink{0000-0002-0581-7154}}
\author[b]{B.~Pire\,\orcidlink{0000-0003-4882-7800}}
\author[a]{P.~Sznajder\,\orcidlink{0000-0002-2684-803X}}
\author[a]{J.~Wagner\,\orcidlink{0000-0001-8335-7096}}

\affiliation[a]{National Centre for Nuclear Research (NCBJ),\\
  02-093 Warsaw, Poland}

\affiliation[b]{Centre de Physique Th\'eorique, CNRS, \'Ecole Polytechnique,\\
I.P. Paris, 91128 Palaiseau, France}

\emailAdd{victor.martinez-fernandez@ncbj.gov.pl}
\emailAdd{bernard.pire@polytechnique.edu}
\emailAdd{pawel.sznajder@ncbj.gov.pl}
\emailAdd{jakub.wagner@ncbj.gov.pl}

\abstract{Generalized parton distributions (GPDs) are off-forward matrix elements of quark and gluon operators that work as a window to the total angular momentum of partons and their transverse imaging (nucleon tomography). To access GPDs one needs to look into exclusive processes which are usually studied in a kinematic regime known as the Bj\"orken limit. In this limit, the photon virtualities are much larger than the hadron mass $M$, and the kick to the hadron measured by the Mandelstam's variable $t$. It turns out that this is not enough for the purposes of a precise GPD extraction and, in particular, of nucleon tomography for which measurements in a sizable range of $t$ are required. Deviation with respect to the Bj\"orken limit induces kinematic higher-twist corrections which enter the amplitudes with powers of $|t|/\scale^2$ and $M^2/\scale^2$, where $\scale^2$ denotes the scale of the process (basically, the sum of photon virtualities in the case of DDVCS). There are also corrections by the name of ``genuine'' higher twists which are a separate topic and are not the subject of this research study.

In this manuscript, we present novel calculations of DDVCS amplitudes off a (pseudo-)scalar target including up to kinematic twist-4 corrections. These results are important for measuring DDVCS, DVCS and TCS through the Sullivan process and off helium-4 target at the future Electron-Ion Collider (EIC) and JLab experiments. Preliminary numerical estimates for the pion target are provided.}

\FullConference{31st International Workshop on Deep Inelastic Scattering (DIS2024)\\
 8–12 April 2024\\
Grenoble, France\\}

\begin{document}
\maketitle

\section{Introduction}\label{sect::intro}

Generalized parton distributions (GPDs) are off-forward matrix elements of quark and gluon operators that appear in the scattering amplitudes for deep exclusive processes. They constitute a 3D version of the usual one-dimensional parton distribution function (PDF). For a quark of flavor $f$ in a spin-0 hadron, there is only one chiral-even leading-twist GPD that is defined as:
\begin{equation}
    H_f(x,\xi,t) = \int \frac{dz^-}{2\pi} e^{ix\bar{p}^+z^-} \langle p'|\bar{\quark}_f(-z/2)\slashed{n}{\cal W}[-z/2, z/2] \quark_f(z/2) |p\rangle\big|_{z_\perp = z^+ = 0}\,,
\end{equation}
where ${\cal W}$ represents a Wilson line, $\xi=-\Delta n/(2\bp n)$ is the so-called skewness and the vector $n$ is the usual lightlike vector projecting the positive-longitudinal component in light-cone coordinates. Also, $\bp=(p+p')/2$.

\begin{figure}
    \centering
    \includegraphics[scale=0.7]{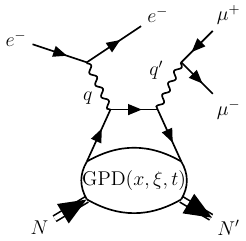}
    \caption{\scriptsize{Double deeply virtual Compton scattering (DDVCS) amplitude at lowest order in the muon channel.}}
    \label{fig::ddvcs}
\end{figure}

Double deeply virtual Compton scattering (DDVCS, illustrated in Fig.~\ref{fig::ddvcs})~\cite{PhysRevLett.90.022001,Belitsky:2003fj,guidal2003,Deja:2023ahc}:
\begin{equation}
    \gamma^*(q) + N(p)  \to \gamma^*(q') + N'(p') \,,
    \label{reaction2}
\end{equation}
contributes to the exclusive electroproduction of a lepton pair off a hadron $N$, which in the muon channel is given by the reaction
\begin{equation}
    e(k) + N(p)  \to e'(k') + N'(p') + \mu^+(\ell_+) + \mu^-(\ell_-) \,.
    \label{reaction}
\end{equation}
This process also receives contributions from the purely QED Bethe-Heitler processes (BH), which do not provide access to GPDs. Hence, they are not affected by the twist expansion and will be ignored in what follows. See Refs.~\cite{Belitsky:2003fj,guidal2003,Deja:2023ahc} for details on BH and its interplay with DDVCS regarding the cross-section and observables of the reaction~(\ref{reaction}) for a spin-1/2 target $N$.

Other related processes are deeply virtual Compton scattering (DVCS)~\cite{Muller:1994ses,Ji:1996nm,Radyushkin:1998bz} and timelike Compton scattering (TCS)~\cite{Berger:2001xd}. These two can be described under the formalism of DDVCS: taking the appropriate small virtuality limit, one recovers DVCS and TCS from DDVCS results~\cite{Deja:2023ahc}.
DDVCS factorizes by means of GPDs and perturbatively calculable hard-coefficient functions when the scale of the process $\scale^2$ (basically, the sum of photon virtualities, $Q^2+Q'^2$ with $Q^2=-q^2$ and $Q'^2=q'^2$) is large enough: $|t|/\scale^2,\, M^2/\scale^2\to 0$. This kinematic leading twist (LT) approximation is then applied to the Compton tensor
\begin{equation}\label{comptontensor}
    T^{\mu\nu}= i\int d^4z\ e^{i\bq z}\langle p'| 
    \mathbb{T}\left\{ j_\mu(z/2) j_\nu(-z/2) \right\} |p\rangle\,,\quad \bq=\frac{q+q'}{2} \,,
\end{equation}
to get the LT DDVCS amplitude where GPDs enter by means of Compton form factors (CFFs) defined, at leading order (LO) in $\alpha_s$, as
\begin{equation}\label{cffHddvcs}
    \cffH(\rho,\xi,t) = -\textrm{PV}\int_{-1}^1dx\ \frac{1}{x-\rho}H^{(+)}(x,\xi,t) + i\pi H^{(+)}(\rho,\xi,t) \,,
\end{equation}
where $\hplus$ is  the C-even part of the GPD $H$ weighted by the squared fractional quark electric charges:
\begin{equation}
    H^{(+)}(x,\xi,t) = \sum_f \left(\frac{e_f}{e}\right)^2 \left( H_f(x,\xi,t) - H_f(-x,\xi,t) \right)\,,
\end{equation}
Also, in Eq.~(\ref{cffHddvcs}), the generalized Bj\"orken variable $\rho$ has been introduced as $\rho=\xi (qq')/(\Delta q')$.

Current and future experimental data cannot be considered as a trustworthy realization of the $|t|/\scale^2\to 0$ and $\, M^2/\scale^2\to 0$ limits. Therefore, corrections accounting for these effects need to be included and are referred to as kinematic higher-twist corrections, involving the leading-twist GPDs. There are also ``genuine'' higher twist corrections involving new GPDs, which are a separate topic and are not considered here. One can relate the kinematic corrections to the deviations with respect to the light-cone dominance of the Compton tensor~(\ref{comptontensor}). With this idea in mind and taking into account that at LO QCD is a conformal field theory (CFT),\footnote{Beyond LO, QCD is a conformal field theory at the Fisher-Wilson fixed point of the $\beta-$function. Working at said point and using the renormalization group equations, one can obtain correct results for QCD at four dimensions beyond LO. For further details, cf.~\cite{Braun:2016qlg} for example.} V.~M.~Braun et al.~developed an all kinematic-twist expansion of said tensor in Refs.~\cite{Braun:2020zjm, Braun:2022qly}. This methodology is briefly described in the next section.

\section{Conformal operator-product expansion and the scale of DDVCS}\label{sect::cope+scaleDDVCS}

Making use of the shadow-operator formalism of CFTs, cf.~\cite{ferrara1972}, one can obtain a close expression for the LO Compton tensor valid to all kinematic twists. For illustration purposes, some of the most light-cone divergent terms are given by
\begin{align}\label{comptontensor_COPE}
     T^{\mu\nu} = &\ i\int d^4z\ e^{iq'z}\langle p'| \mathbb{T}\{j^\mu(z)j^\nu(0)\} |p\rangle = \frac{1}{i\pi^2}i\int d^4z\ e^{iq'z}\Bigg\{ \frac{1}{(-z^2+i0)^2}\Bigg[ 
    g^{\mu\nu}\lop(1,0) \nonumber\\
    & \qquad\qquad - z^\mu\partial^\nu\int_0^1 du\ \lop(\bu, 0) - z^\nu(\partial^\mu - i\Delta^\mu)\int_0^1 dv\ \lop(1,v) \Bigg] + \cdots\,,
\end{align}
where symbols $\lop$ above are to be understood as matrix elements of the corresponding operators,~i.e.
\begin{equation}\label{lopMatrixElement}
    \langle p'| \lop(\lambda_1, \lambda_2) |p\rangle = \frac{2i}{\lambda_{12}}\int_{-1}^1 d\beta\int_{|\beta|-1}^{1-|\beta|} d\alpha\ \lt{e^{-i\ell_{\lambda_1,\lambda_2}z}}\Phi^{(+)}(\beta,\alpha,t)\,,\quad \lambda_{12}=\lambda_1-\lambda_2\,,
\end{equation}
and
\begin{equation}
    \ell_{\lambda_1,\lambda_2} = -\lambda_1\Delta - \lambda_{12}\left[ \beta\bp - \frac{1}{2}(\alpha+1)\Delta \right]\,,\quad \bp=\frac{p+p'}{2}\,,\quad \Delta = p'-p\,,
\end{equation}
for $\alpha,\beta$ the parameters of the usual double distributions (DDs) $h_f(\beta,\alpha,t), g_f(\beta,\alpha,t)$~\cite{Radyushkin:1998bz}, which are related to the function $\phiplus$ as
\begin{equation}
    \PhiPlus = \sum_f \left(\frac{e_f}{e}\right)^2\PhiPlusF\,,\quad \Phi_f^{(+)}(\beta,\alpha,t) = \partial_\beta h_f + \partial_\alpha g_f \,.
\end{equation}
By $\lt{\ }$ we refer to the {\it geometric} LT projection which reduces an operator to its symmetric and traceless component. Such a projector is described in Ref.~\cite{geyer1999}.

After performing the Fourier transform of the different components of the geometric LT projection of the exponential in (\ref{lopMatrixElement}), one is left with integrals of the form
\begin{equation}
    I_{n,m}=\int_0^1 dw\ \frac{w^n}{(aw^2+bw+c)^m}\,,\quad a = \ell_{\lambda_1,\lambda_2}^2\,,\quad b = -2q'\ell_{\lambda_1,\lambda_2}\,,\quad c=Q'^2+i0\,.
\end{equation}
The structure of these integrals suggests the scale for any two-photon process to be $\scale^2=-2q' \Delta=Q^2+Q'^2+t$, as $a \propto (|t|,\,M^2)$ and $b \propto -2q'\Delta\approx Q^2+Q'^2 > 0$. The $t$ factor in $\scale^2$ could be dropped as it produces corrections of the next twist. Therefore, the twist expansion comes in powers of $a/b = O(\textrm{tw-4})$, whereas $c/b = O(1)$.

In the next section, we apply these techniques to DDVCS off a spin-0 target.

\section{Scalar and pseudo-scalar target}\label{sect::(pseudo-)scalar}

The calculation of twist corrections can be organized by means of the so-called helicity-dependent amplitudes $\amp^{AB}$ which are the amplitudes for transition between an incoming photon with helicity $A$ to an outgoing photon with helicity $B$, where $A,B\in\{0,\pm 1\}$. The (helicity-dependent) CFFs of the hadron are related to them by $\amp^{AB} = \cffH^{AB}/2$. Therefore, the Compton tensor can be parameterized in terms of these amplitudes through:
\begin{align}
    T^{\mu\nu} = &\ \amp^{00} \frac{-i}{QQ'R^2}\left[ (qq')(Q'^2q^\mu q^\nu - Q^2 q'^\mu q'^\nu) + Q^2 Q'^2 q^\mu q'^\nu - (qq')^2 q'^\mu q^\nu \right] \nonumber\\
    & + \amp^{+0}\frac{i\sqrt{2}}{R|\bp_\perp|}\left[ Q'q^\mu - \frac{qq'}{Q'}q'^\mu \right]\bp^{\,\nu}_\perp -\amp^{0+}\frac{\sqrt{2}}{R|\bp_\perp|}\bp^{\,\mu}_\perp\left[ \frac{qq'}{Q}q^\nu + Q q'^\nu \right] \nonumber\\
    & + \amp^{+-}\frac{1}{|\bp_\perp|^2}\left[ \bp^{\,\mu}_\perp\bp^{\,\nu}_\perp - \widetilde{\bp}^{\,\mu}_\perp\widetilde{\bp}^{\,\nu}_\perp \right] - \amp^{++}g_\perp^{\mu\nu}\,,
\end{align}
where $R = \sqrt{(qq')^2 + Q^2Q'^2}$, $\widetilde{\bp}^\mu_\perp=\epsilon^{\mu\nu}_\perp \bp^\nu$ and $\amp^{AB}=\amp^{-A\,-B}$ due to parity conservation. From this expression, a set of projectors onto the different $\amp^{AB}$ can be read out and applied to Eq.~(\ref{comptontensor_COPE}). Let us now present $\amp^{+-}$ as an example, together with preliminary results on its magnitude.

\subsection{Example: transverse-helicity flip amplitude, $\amp^{+-}$}\label{sect::A+-}

The transverse-helicity flip amplitude ($\amp^{+-}$) is of a special interest as it may contribute to the cross-section in two ways: i) at LO as a higher-twist contribution, and ii) at NLO via gluon transversity GPDs~\cite{Belitsky:2000jk}. We stay at LO, therefore $\amp^{+-}$ starts at kinematic twist-4 and we obtain
\begin{align}
    \amp^{+-}= &\ \frac{1}{2|\bp_\perp|^2}\left( \bp_{\perp,\,\mu}\bp_{\perp,\,\nu} - \widetilde{\bp}_{\perp,\,\mu}\widetilde{\bp}_{\perp,\,\nu} \right) T^{\mu\nu} = \frac{4\bp_\perp^2}{\scale^2} D_\xi^2 \int_{-1}^{1} \frac{dx}{2\xi}Y\left(\frac{x}{\xi},\frac{\rho}{\xi}\right)~H^{(+)}(x,\xi, t)\,, 
\end{align}
where $D_\xi = \xi^2\dxi$, $\dxi = \partial/\partial\xi$ and 
\begin{equation}
    Y\left(\frac{x}{\xi},\frac{\rho}{\xi}\right) = -\frac{\xi + \rho}{\xi + x} \log \frac{\rho - x -i0}{\xi + \rho} -\frac{\xi - \rho}{\xi - x} \log \frac{x-\rho +i0}{\xi - \rho} +2 \log \frac{\rho -x -i0}{2\xi}\,,
\end{equation}
for which the DVCS and TCS limits take the form:
\begin{equation}
     Y_{\rm DVCS}\left(\frac{x}{\xi}\right) = \frac{2x}{\xi + x} \log \frac{\xi - x -i0}{2\xi } \,,\quad  Y_{\rm TCS}\left(\frac{x}{\xi}\right) = \frac{2x}{ x -\xi} \log \frac{ \xi + x +i0}{2\xi }\,.
\end{equation}
Our calculation agrees in the DVCS limit with previous works~\cite{Braun:2022qly} and with the DVCS-TCS connection discussed in~\cite{Mueller:2012sma}.

\subsection{Numerical estimates of $\amp^{+-}$}

\begin{figure}[ht]
    \centering
    \includegraphics[scale=0.3]{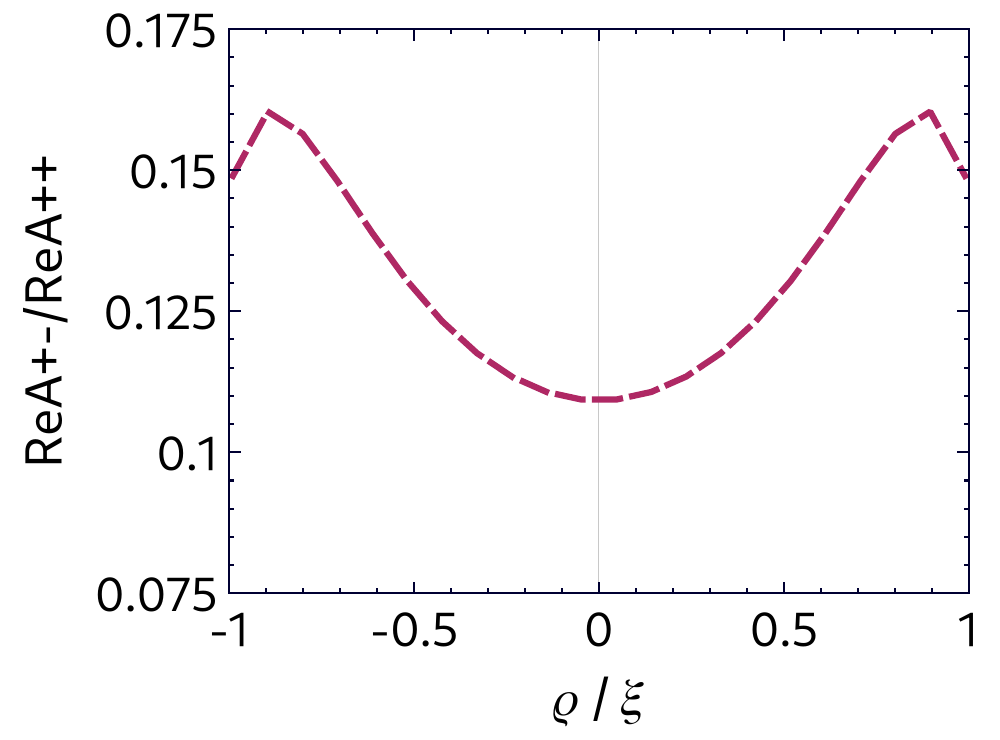}~~~~~~~~
    \includegraphics[scale=0.3]{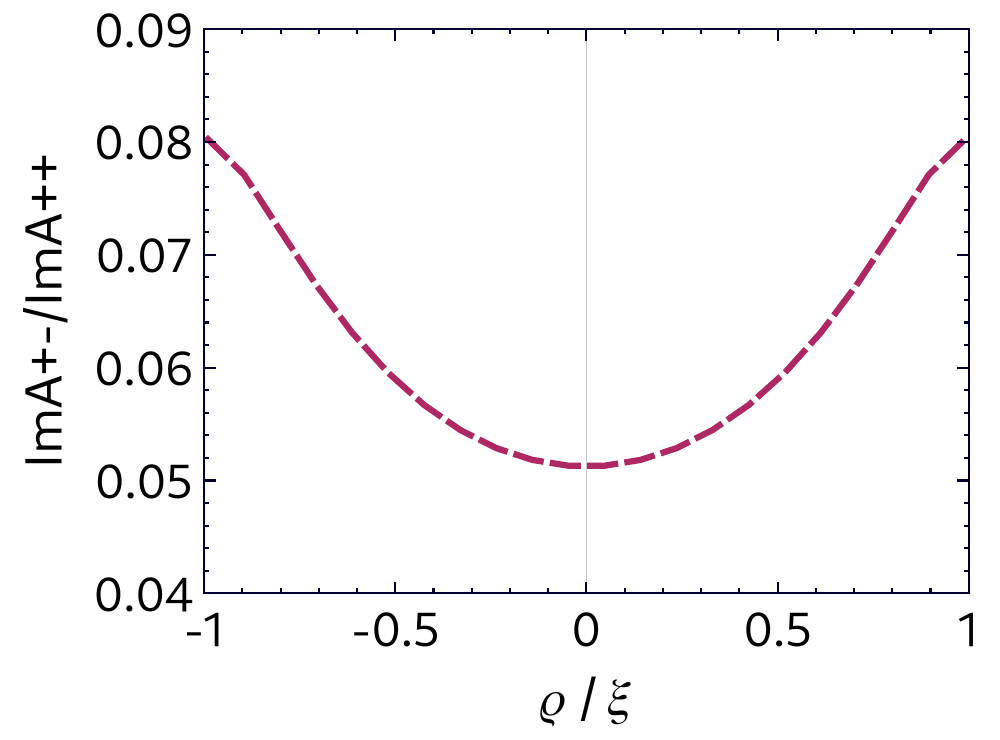}
    \caption{\scriptsize{Real (left) and imaginary (right) parts of $\amp^{+-}$ relative to $\left.\amp^{++}\right|_{\textrm{LT}}$ as a function of $\rho/\xi$ for $\xi = 0.25$, $\scale^2 = 1.9~\mathrm{GeV}^2$ and $t = -0.8~\mathrm{GeV}^2$.}}
    \label{fig::A+-_vs_rhoOverXi}
\end{figure}

We compare in Fig.~\ref{fig::A+-_vs_rhoOverXi} the numerical estimate of the real and imaginary parts of the amplitude $\amp^{+-}$  with the leading-twist contribution to the Compton tensor, which comes solely from the transverse-helicity conserving amplitude $\amp^{++}$. We use the phenomenological pion-GPD model described in Ref.~\cite{PhysRevD.105.094012}. The extreme left ($\rho/\xi=-1)$ and right ($\rho/\xi=+1$) values of the horizontal axes represent the TCS and DVCS limits, respectively. We observe that $\amp^{+-}$ is of the order of $\sim 10\%$ with respect to $\left.\amp^{++}\right|_{\textrm{LT}}$ for \mbox{$-t/\scale^2 \approx 0.5$}, which should be a measurable effect in DVCS as well as future TCS experiments.

\noindent\small{{\bf Acknowledgements.} The works of V.M.F.~are supported by PRELUDIUM grant 2021/41/N/ST2/00310 of the Polish National Science Centre (NCN).}

\scriptsize
\bibliographystyle{JHEP}
\bibliography{biblio}

\end{document}